\documentclass{iucr}             
\usepackage {graphicx}
\usepackage {amsmath}
\usepackage {bm}

    \paperprodcode{a000000}      
     \paperref{xx9999}            
     \papertype{FA}               

     \paperlang{english}          
     \journalcode{A}              
     \journalyr{2004}
     \journaliss{0}
     \journalvol{60}
     \journalfirstpage{000}
     \journallastpage{000}
     \journalreceived{0 XXXXXXX 0000}
     \journalaccepted{0 XXXXXXX 0000}
     \journalonline{0 XXXXXXX 0000}

\begin{document}                  



\shorttitle{SIR2002 and the non-crystallographic phase problem}
\title{Phasing diffuse scattering. Application of the SIR2002 algorithm to 
the non-crystallographic phase problem.}


\author[a]{B.}{Carrozzini} 
\author[a]{G. L.}{Cascarano}
\author[a]{L.} {De Caro}
\cauthor[a]{C.} {Giacovazzo}{Email: carmelo.giacovazzo@ic.cnr.it}

\author[b]{S.}{Marchesini}
\author[b]{H.~N.} {Chapman}

\author[c]{H.} {He}
\author[c]{M.~R.} {Howells} 
\author[d]{J.~S.}{Wu} 
\author[d]{U.} {Weierstall}
\author[d]{J.~C.~H.}{Spence}

\shortauthor{B.~Carrozzini, G.~L.~Cascarano,~L.~De Caro et al.}

\aff[a]{Campus Universitario, Via E. Orabona 4, 70125, Bari \country{Italy}}
\aff[b]{Lawrence Livermore Laboratory, 7000 East Ave., Livermore, Ca 94550-9234, \country{USA}}
\aff[c]{Lawrence Berkeley Laboratory, 1 Cyclotron Rd. Berkeley, Ca., 94720, \country{USA}}

\aff[d]{Dept. of Physics and Astronomy, Arizona State University, Tempe, Az. 85287-1504. \country{USA}}










\maketitle                        

\begin{synopsis}
A new algorithm for determining the phases of 
diffuse elastic X-ray scattering from a non-periodic object is described.
\end{synopsis}

\begin{abstract}
A new phasing algorithm has been used to determine the phases of diffuse 
elastic X-ray scattering from a non-periodic array of gold balls of 50 nm 
diameter. Two-dimensional real-space images , showing the charge-density 
distribution of the balls, have been reconstructed at 50 nm resolution from 
transmission diffraction patterns recorded at 550 eV energy. The 
reconstructed image fits well with scanning electron microscope (SEM) image 
of the same sample. The algorithm, which uses only the density modification 
portion of the SIR2002 program, is compared with the results obtained \textit{via} the 
Gerchberg-Saxton-Fienup HIO algorithm. In this way the relationship between 
density modification in crystallography and the HIO algorithm used in signal 
and image processing is elucidated.
\end{abstract}


\section{Introduction}

The phase problem for non-periodic objects (Stark, 1987) has been studied by 
many different approaches, some of which have recently demonstrated striking 
success with experimental data:

\begin{enumerate}
\renewcommand{\theenumi}{\alph{enumi}}
\item
the Gerchberg-Saxton-Fienup HIO algorithm (Fienup, 1982, 1987; Miao \textit{et al.}, 1999, 
2002, who refer the method as \textit{oversampling; }see also De Caro\textit{ et al., }2002 who refer to it by using 
the concept of\textit{ confined structure});
\item
techniques based on analyticity and complex zeros (Liao \textit{et al}., 1997);
\item
the study of projections onto convex sets (Bauschke \textit{et al}., 2002 );
\item
the transport of intensity equations (Paganin {\&} Nugent, 1998);
\item
direct methods, for real and positive objects (Spence, \textit{et al}., 2003).
\end{enumerate}
However experimental implementation of the popular HIO algorithm in the 
transmission geometry is complicated by loss of low spatial frequencies at 
the beam-stop. To overcome this problem, in the past the HIO oversampling 
method has been used in combination with an independent low-resolution image 
of the object ($i.e.$ from SEM, X-ray zone-plate or optical image) to provide the 
low spatial frequencies. More recently, the development of the ``Shrinkwrap" 
(Marchesini \textit{et al.}, 2003) and other variants of the HIO algorithm have 
successfully treated the missing data as adjustable parameters. 
Experimentally, the preparation of the isolated object needed for HIO can be 
difficult; only recently the first atomic resolution image of a carbon 
nanotube has been reconstructed by HIO from a coherent transmission electron 
diffraction pattern using nanotubes which span holes in a carbon grid (Zuo 
\textit{et al}., 2003).

In a recent paper (He \textit{et al.}, 2003) images of randomly placed two-dimensional arrays of gold balls were reconstructed from their soft X-ray transmission 
diffraction patterns by means of the HIO algorithm. This algorithm is 
closely related to the electron density modification procedure of X-ray 
crystallography, and is able to: i) exploit the prior knowledge of the 
geometrical form of a support with vanishing electron density (the support 
is defined as the region in which the object density is known to be 
non-zero: it corresponds roughly to the boundary of the object); ii) treat, 
\textit{via} suitable constraints, the real and imaginary parts of the object 
wavefunction (Miao \textit{et al}., 1999). A necessary tool for success was the use of 
either a mask based on supplementary images provided by SEM techniques, or a 
subjective estimate of the object boundary taken from the autocorrelation 
function. 
In the ``Shrinkwrap" method (Marchesini \textit{et al}., 2003), 
a first estimate 
of the object support is obtained from the autocorrelation function of the 
object, (which is known to be twice as large as the object in any given 
direction), and this estimate is updated iteratively. If the structure is 
constituted by one isolated ball and two clusters (as in the case treated 
here) the autocorrelation function includes an image of every cluster 
convoluted with the single ball, and this may used as a trial support 
function. The method is then similar to the heavy-atom method of X-ray 
crystallography, or Fourier Transform Holography. 

In a companion paper (Spence \textit{et al}., 2003), direct methods 
(DM) have been applied 
to a simulated diffraction pattern calculated from a structure of 28 gold 
balls. In that work the phasing process was realized by the following steps. 
The simulated pattern was divided by the Fourier transform $f(\bm{h})$ 
of a single ball, so obtaining a simulated normalized pattern (SNP). The 
structure factors, in turn, were also normalized, according to 

\begin{equation}
\label{eq1}
E_{h} = \sum\limits_{j = 1}^{N} {f_{j} (h)\exp \left( 2\pi ih \cdot r_{j}  
\right) / \left[ \sum\limits_{j = 1}^{N} f_{j} (h)^{2}  \right]} ^{1 / 
2},
\end{equation}
where $N$ is the total number of balls. Then a modified version of the SIR2002 
program (Burla \textit{et al.}, 2003; a package originally dedicated to the crystal 
structure solution of small and large molecules) was used to determine the 
positions of the balls from the SNP. The procedure consisted of a tangent 
procedure starting from random phases, followed by:

\begin{enumerate}
\renewcommand{\theenumi}{\alph{enumi}}
\item
 the EDM step, where new phase sets $\{\varphi \}$ were estimated by 
iteratively applying electron density modification procedures, i.e., $\rho 
\to \left\{ \varphi \right\} \to \rho$ cycles;
\item
 the HAFR step, consisting of $\rho \to \left\{ \varphi \right\} \to \rho 
$ cycles, in which balls were associated with electron density peaks;
\item
 Least squares cycles alternating with ($2F_{obs}-F_{calc}$) map 
calculations to refine and complete the structural model.
\end{enumerate}

Such an approach requires prior knowledge of the scattering factors of the 
balls and of their number $N.$ However it does not need prior knowledge of the 
geometry of the support with vanishing electron density. When applied to the 
experimental speckle pattern considered in this paper, such an algorithm did 
not succeed, probably owing to unavoidable errors in the measurements and in 
the model scattering function (the gold balls do not exactly have the same 
radius). 

This negative result has stimulated the search of a more general phasing 
procedure, described in this paper, based on the following general 
conditions:

\begin{enumerate}
\renewcommand{\theenumi}{\alph{enumi}}
\item
no prior knowledge of the scattering factor of the gold balls or of their 
number; 
\item
\noindent
no use of any support, as masks based on the autocorrelation function or SEM 
images.
\end{enumerate}

While a diffuse scattering pattern may be generated by a complex 
transmission function in real space, the algorithm described in this paper 
is restricted to real densities, such as the charge density for X-rays, or 
the electrostatic potential for electron diffraction. Examples of complex 
objects include those described by a complex refractive index with 
significant spatial variation in absorption, or multiple scattering effects 
in electron or X-ray diffraction. We then assume that Friedel's law holds, 
and note that this constraint may hinder the precise reconstruction of the 
structure, if the imaginary component of the scattering is not negligible 
with respect to the real part.

\section{Experimental}

\begin{figure}
\includegraphics[width=7cm]{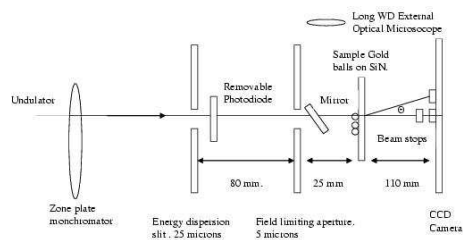}
\caption{Set up of the diffraction experiment. \label{fig1}}
\end{figure}

The experimental soft X-ray transmission diffraction pattern from two 
clusters of gold balls of $50 \pm  5$ nm diameter deposited on a silicon 
nitride window was recorded at the Advanced Light Source at the Lawrence 
Berkeley Laboratory, using soft x-rays with a wavelength of 2.1 nm (He \textit{et al.}, 
2003). The experiments used the ``pink'' 588 eV (2.11 nm) photon beam at 
beam-line 9.0.1. Features of the beam line include: a 0.5 $\mu $m-thick, 750 
$\mu $m-square Be window to separate the UHV beam line from the low-vacuum 
sample environment; a monochromator consisting of an off-axis segment of a 
zone plate and the diffraction experiment itself ( Fig. \ref{fig1}). Samples were 
made by placing a droplet of solution containing `gold conjugate' colloidal 
gold balls on a silicon nitride window (thickness 100 nm) and allowing it to 
dry. The gold balls formed several single layered (2D) clusters on the SiN 
membranes, as imaged by a field-emission scanning electron microscope (SEM).

The x-ray coherence length $l_{c}$ must be greater than the maximum path 
difference between any pair of interfering rays, i.e. $l_{c}>w \Theta_{max}$ where 
$w$ is the width of the sample and $\Theta_{max}$ is the maximum diffraction angle. 
For our geometry and wavelength, $\Theta_{max} = 0.12$ radians and the resolution 
limit is 8.4 nm. For the 5$\mu $m aperture (effectively the monochromator 
exit slit) shown in Fig. \ref{fig1}, the resolving power is about 500, the coherence 
length is then 1$\mu$m and the maximum sample illumination area 8$\times $8 
$\mu$m$^{2}$. Similarly the (spatial) coherence patch provided by the 5 
$\mu$ m aperture is 10$\times $10 $\mu $m$^{2}$. Allowing for an empty (but 
still coherently illuminated) band around the sample, its allowed size is 
thus $<4\times $4 $\mu $m$^{2}$. 

The periodic boundary conditions which are necessarily imposed in the DM 
approach are now considered. Correct sampling of the diffraction-plane 
intensity avoids aliasing of the autocorrelation function, and is given by 
Shannon's sampling theory. The Shannon interval for frequency-space sampling 
of the \textit{intensity} is 1/(2w)=$\Delta $/$\lambda $z where $z$ is the sample-to-detector 
distance and $\Delta $ is the detector-plane increment (a 25 $\mu $m CCD 
pixel in our case). For our $\Delta $ and $z$ values the Shannon theorem also 
leads to a maximum sampling interval of 4 $\mu $m. This is the correct 
sampling of the diffraction-plane \textit{intensity} involving a twofold \textit{over}sampling in each 
direction of the diffraction-plane wave complex amplitude. 

In place of the atomic scattering factor required by DM, we use the 
scattering distribution for a gold ball at 550 eV. The semiangle \textit{$\Theta $ } (see 
Fig. \ref{fig1}) subtended at the object by the first detector pixel beside the optic 
axis defines the lowest spatial frequency $1/D= \Theta /\lambda $ 
contributing to the reconstruction, and we treat this pixel as the 
first-order Bragg reflection - (1 0) Miller indices - for a square 
two-dimensional unit cell with symmetry P1. The use of SIR2000 imposes 
periodic continuation of this cell. For the performed experiments, $D = 9.147$ 
$\mu $m. For our $1025\times 1025$ CCD, the highest order reflection is 
(512,512); the real-space resolution for pixel (512,0) at the boundary is 
D/512 = 17.8 nm, giving about four pixels per gold ball in the reconstructed 
image. The width of the region illuminated by the beam is about $B=5$ $\mu 
$m, which is less than the coherence width of the beam. The diffraction 
pattern consists of the Airy's disk-like pattern from a single ball, crossed 
by finer ``speckle" fringes due to interference between different balls. The 
first zero for a ball rather than a disk occurs at a semiangle 1.4 $\lambda 
$/$d$, which occurs at the reflection with pixel (Miller) indices (256, 0). 
(The spacing D1 corresponding to the angle 1.4$\times \lambda $/d is: 
D1=d/1.4=50/1.4=35.7 nm. As we have assumed D=9.147 micron, the index is: 
D$\times $1.4/50=256).

\begin{figure}
\includegraphics[width=7cm]{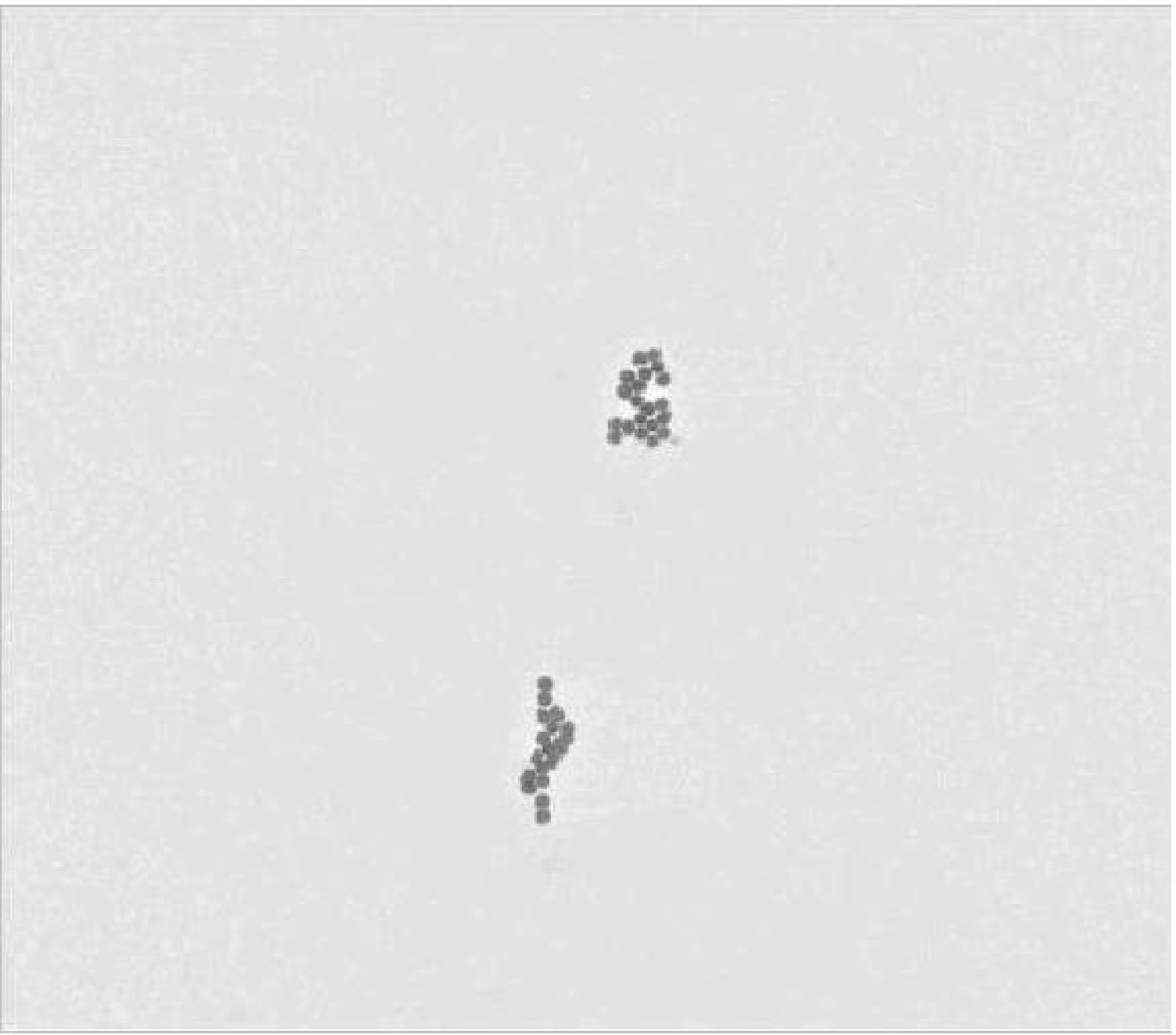}
\caption{SEM image of a random set of gold balls of 50 nm diameter 
at 550 eV. \label{fig2}}
\end{figure}

\begin{figure}
\includegraphics[width=7cm]{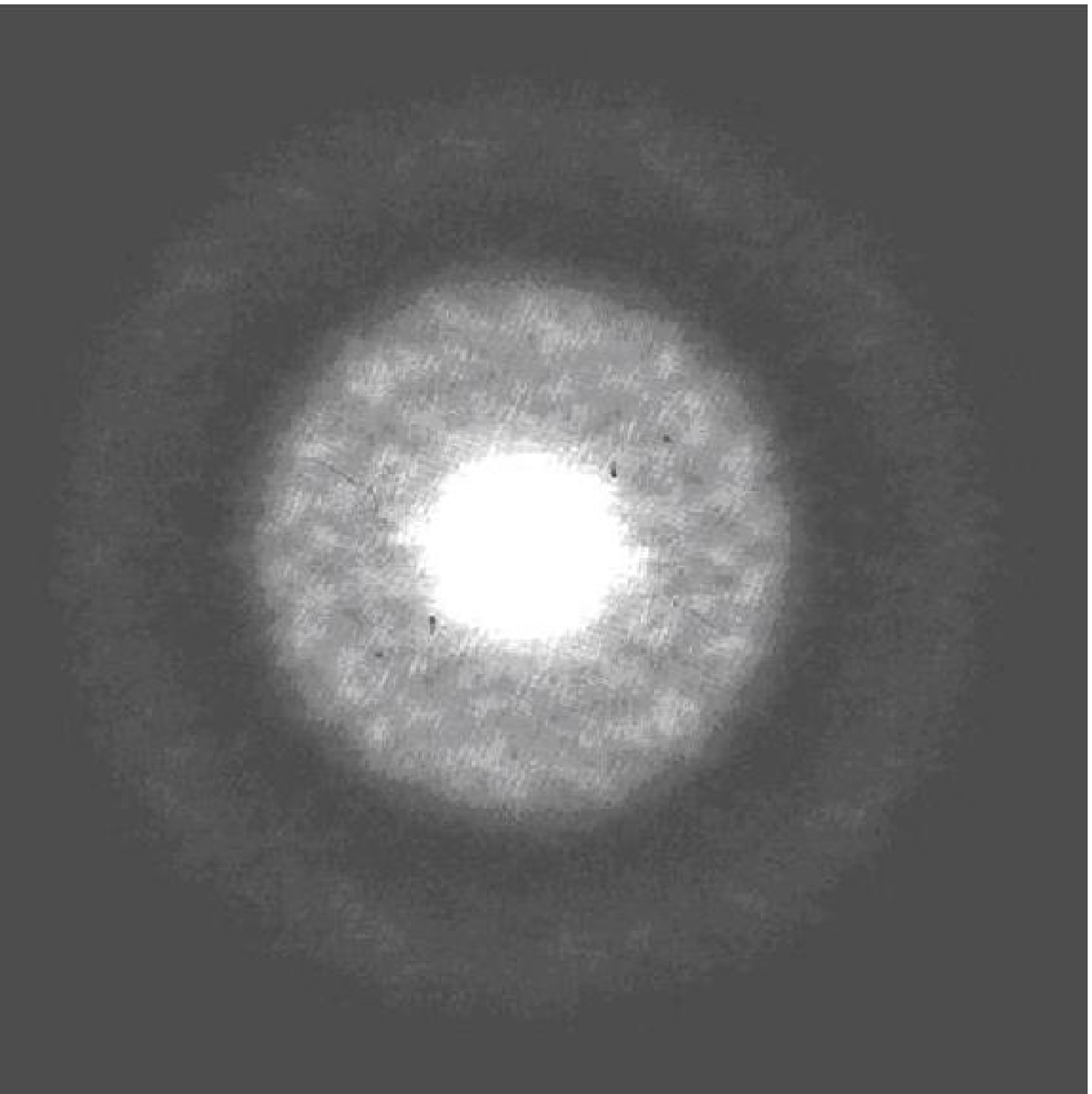}
\caption{Experimental soft X-ray transmission diffraction pattern 
from the 50nm diameter gold balls shown in Fig.2. The X-ray wavelength is 
2.5 nm. \label{fig3}}
\end{figure}

Figure \ref{fig2} shows a SEM image of a random set of gold balls of 50nm diameter at 
550 eV. Figure \ref{fig3} gives the corresponding diffraction pattern, showing fine 
speckle fringes modulating the pattern expected from a sphere. The first 
minimum of this pattern occurs at a value of sin$\theta $/$\lambda $ = 
1.4/$d$, where $d$ is the diameter of the ball (the factor 1.4 is replaced by 1.2 
for a disk). 

The number of the measured reflections (Friedel pairs included) is 992061, 
for data resolution up to 126 {\AA}, about one quarter of the gold ball 
diameter; the number of measured unique reflections (Friedel opposite 
merged) is REFL = 522517; the number of unique reflections, not measured 
because of the beam stop, is UREFL = 1365.

\section{The new phasing procedure }

The new procedure uses only the electron density modification subroutine of 
the SIR2002 software package. In real space, only a small portion of pixels 
with the highest values (about 0.25 percent) were selected as defining the 
region inside the support for the current estimate of the density function. 
In reciprocal space, the phasing proceeded gradually from low-order 
reflections outward. The modulus constraint was applied in each cycle. The 
initial assignment of phases is random. We assume that:

\begin{enumerate}
\renewcommand{\theenumi}{\alph{enumi}}
\item
the nature and the number of the scatterers are unknown (consequently, their 
scattering factors are also unknown);
\item
anomalous dispersion effects can be neglected. Accordingly, only the set of 
unique reflections will be considered in the calculations. Thus the 
intensities of the Friedel opposites are merged.
\end{enumerate}

The above assumptions exclude most of the steps of the standard SIR2002 
procedure: e.g., the normalization of the structure factors , the tangent 
process, the HAFR refinement cycles and the least squares process (see 
section 1). Accordingly, the new SIR2002 procedure consists only of cycles 
of electron density modification $\rho \to \left\{ \varphi \right\} \to \rho 
$. The phasing process may be partitioned into two steps.

STEP 1 - Because of their importance (see section 4), only low resolution 
reflections are used initially in the phasing process. A resolution 
threshold TRH is defined, equal to 0.25 of the experimental data resolution. 
The program starts by assigning random phases to a very limited number 
(2500) of reflections. Then 60 macrocycles, each constituted by 9 
microcycles $\rho \to \left\{ \varphi \right\} \to \rho $ are performed. 
During a macrocycle the number of phased reflections used in each hemicycle 
$\left\{ \varphi \right\} \to \rho $ progressively increases up to NREFTRH 
(this is the number of unique reflections satisfying the threshold 
resolution TRH; in our case NREFTRH= 52500). In each hemicycle $\rho \to 
\left\{ \varphi \right\}$ the number of grid points of the electron density 
map used for calculating phases is 0.0025 of their total number (we select 
the grid values with highest intensity). 

At the end of each macrocycle the program estimates, in the hemicycle $\rho 
\to \left\{ \varphi \right\}$, the intensities and the phases of the UREFL 
reflections. The intensities are used as the observed values in the next 
cycle, and are updated (together with the calculated phases) at the end of 
every macrocycle (modulus constraint).

STEP 2 - The phasing process is extended to the full set of experimentally 
available reflections. The phase values obtained in the STEP 1 for the low 
resolution reflections are used as pivots in the phasing process: their new 
values (obtained at the end of each macrocycle) are always combined with the 
values obtained at the end of the STEP 1. By contrast , the UREFL 
reflections are recalculated as in the STEP 1.

These steps define the relationship between the new proposed algorithm and 
the classical HIO procedure. The new algorithm: 

\begin{enumerate}
\renewcommand{\theenumi}{\alph{enumi}}
\item
varies, from microcycle to microcycle, the number of structure factors used 
in the calculation of $\rho $.

\item
limits the number of grid points in the electron density map, whose 
intensities are used to compute the phases. This practice is \textit{not }suggested by 
the fact that the balls occupy a small part of the total area. Indeed a 
similar procedure is used by the standard version of SIR2002 for solving \textit{ab initio} 
the crystal structure of the proteins, and also of small molecules, which 
fill more densely the unit cell (only about 2{\%} of the pixels are used for 
the electron density inversion). In the case of oversampling, the percentage 
may be further reduced because the unit cell contains a large empty region. 
\end{enumerate}

No information on the geometry of the support is used. This is an advantage 
when such information is unknown or imperfectly defined in the experiment. 

\section{Analysis of the diffraction data and of the results.}

\begin{figure}
\includegraphics[width=7cm]{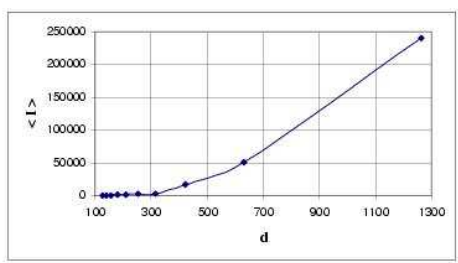}
\caption{The average intensity  $\langle I \rangle$  is plotted against the 
resolution d (in {\AA}). \label{fig4}}
\end{figure}

Two features of the experimental diffraction pattern are evident. The first 
one regards the average intensity $ \langle I \rangle$ as a function of the resolution $d$ 
(calculated in {\AA}), plotted in Fig. \ref{fig4}. We see that the intensities of the 
low order beams are very intense in comparison with the high orders. The 
phasing process, therefore, is expected to be dominated by the inner region 
of the pattern. This is the reason why our phasing procedure tries first: a) 
to phase low frequency reflections, and then uses them as pivots in a phase 
extension process; b) to estimate the number UREFL of unobserved low 
frequency intensities. The subsequent estimates of these beam-stop 
reflections play a crucial role in the success of the procedure: without 
their use the phasing process may fail. A similar observation was reported 
by Wu and Spence (2003) that the intensities of low-resolution fractional 
reflections calculated using a Hilbert transform formula was found to be 
important in performing phase extension. 

The second feature that should be evidenced is that $I(h,k)$ is not perfectly 
symmetrical: significant deviations from Friedel's law were noted. This may 
be due to three reasons: 

\noindent
\begin{enumerate}
\renewcommand{\theenumi}{\alph{enumi}}
\item
 Unavoidable parasitic scattering or absorption. For a well designed 
experiment the effects of parasitic scattering or absorption (border 
effects, silicon nitride absorption, etc\ldots ) should be minimized. The 
success of our phasing process indicates that this is the case for our 
experiment. 

\item
Effects of anomalous scattering. If the balls are assumed to be 
identical, the scattering amplitudes F$_{h,k}$ and F$_{ - h, - k}$ should be 
equal, however the corresponding phases are not constrained to reverse sign 
under index inversion (Spence \textit{et al.} , 2003). Since the radii of the balls satisfy 
a non-delta distribution, and since the balls may be elastically deformed, 
we may be expected that the equality $\vert $F$_{h,k}\vert =\vert 
$F$_{ - h, - k}\vert $ is not strictly obeyed. The deviations, however, 
are expected to be small. 

\item
Incorrect identification of the (0,0) pixel in the detector ( 
\textit{zero-pixel} ) ;
\end{enumerate}

Since the effects of a) are expected to be small, we can identify the 
zero-pixel as the one for which the value of

\begin{equation}
\label{eq2}
R_{merge} = \frac{\sum\nolimits_{h,k} {(\vert F^ + - F_{med} \vert + \vert 
F^ - - F_{med} \vert } )}{\sum\nolimits_{h,k} {(F^ + + F^ - )} }\,,
\end{equation}
is a minimum, where
\begin{equation}
\label{eq3}
F_{med} = \left( {F^ + + F^ - } \right) / 2\,,
\end{equation}
with F$^{ + }$ and F$^{ - }$ to represent F$_{h,k}$ and F$_{ - h, - k}$ 
respectively.

The application of eq. (\ref{eq2}) to the experimental pattern shown in Fig. \ref{fig3} did 
not suggest any origin shift. (We report in the Appendix A the results 
obtained when we applied eq. (\ref{eq2}) to the pattern shown in Fig. \ref{figa1}).

\begin{figure}
\includegraphics[width=7cm]{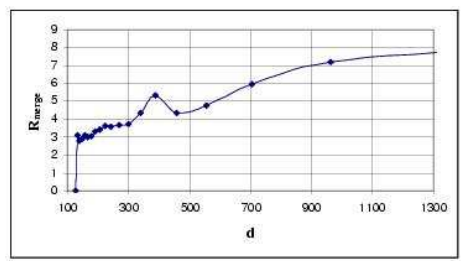}
\caption{$R_{merge} $ versus the resolution d (in {\AA}).\label{fig5}
}
\end{figure}

The deviation from Friedel's law decreases at increasing values of 
sin$\theta $/$\lambda $ (see Fig. \ref{fig5}) where the value of R$_{merge}$ is 
plotted against the resolution d ({\AA}). Such a trend does not comply with 
the expected behaviour of the anomalous dispersion effects, which usually 
increase with sin$\theta $/$\lambda $. This suggests that the deviations 
from Friedel law are mainly due to unavoided parasitic scattering and to 
errors in measurements. (At high resolution, the diffraction intensities are 
particularly faint).

\begin{figure}
\caption{Recovered charge density from non-periodic array of 50nm 
diameter gold balls, using the modified SIR2002 program. \label{fig6}}
\includegraphics[width=7cm]{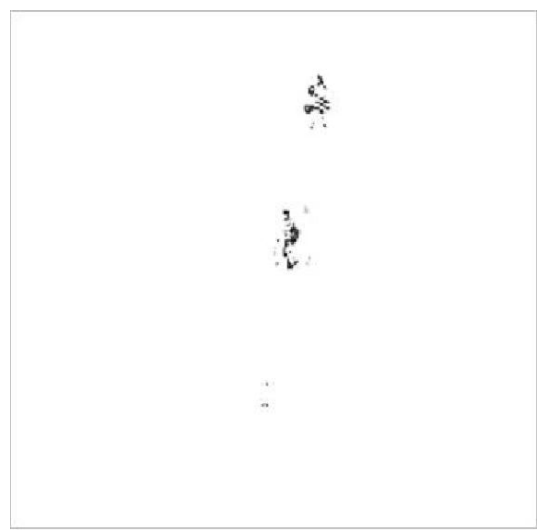}
\end{figure}

We now consider the results of our phasing algorithm. The image resulting at 
the end of the phasing procedure(as shown by the computer screen) is shown 
in Fig. \ref{fig6}: it is in good agreement with the experimental SEM image shown in 
Fig. \ref{fig2} (apart from a shift of origin). 

We require now a quantitative measure of the efficiency of our phasing 
approach, and of the accuracy of the calculated model. Since we have 
ignored the scattering factors of the balls, we cannot use standard 
crystallographic indicators as a measure of agreement between experiment and 
calculations (i.e. the crystallographic residual between the observed and 
the scattered intensities). To overcome this problem we treat the Fourier 
coefficients obtained by Fourier inversion of the final, suitably modified, 
electron density map as the ``calculated" structure factors ($F_{calc}$ in the 
following). The final map produced by SIR2002 is modified (before the 
Fourier inversion) as follows:
\begin{equation}
\rho ( x,y ) =\begin{cases}
 \rho  ( x,y) & \text{if } \rho(x,y) > \langle \rho  \rangle + n \sigma \\
0 & \text{otherwise.}
\end{cases}
\end{equation}
$\langle \rho \rangle$ is the average value of the unmodified electron density map, 
$\sigma $ is its standard deviation. We show in Table \ref{tab1} the residual

\begin{equation}
\label{eq4}
RES = \frac{\sum\nolimits_{h,k} (\left | F_{obs} - 
\vert F_{calc} \vert \right | )} {\sum\nolimits_{h,k} {F_{obs} } }
\,,
\end{equation}

\noindent
for various values of $n$. It may be observed that:

\begin{enumerate}
\renewcommand{\theenumi}{\alph{enumi}}
\item
the value of RES does not significantly change when $n$ varies from 1 to 7. 
For any value of $n$, RES is significantly larger than the figures 
usually obtained for small molecule crystals at the end of the 
structure refinement process. 
Our previous observations suggest that this is probably due to experimental 
errors in the intensities rather than to some intrinsic lack of convergence 
of the phasing algorithm. 
\item
the electron density function is well separated into two domains: the domain 
where the scattering balls are confined, and an empty region;
\item
the refinement process converged to a satisfactory model.
\end{enumerate}

\begin{figure}
\includegraphics[width=7cm]{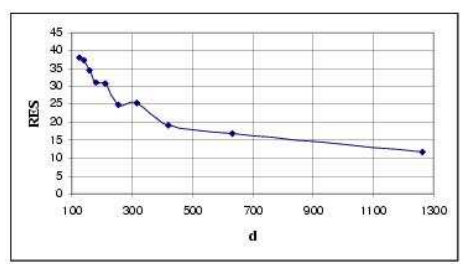}
\caption{Value of RES for $n$ = 1 versus d (in {\AA}). \label{fig7}}
\end{figure}

We have calculated, for the case $n$ = 1, the value of RES against resolution 
(a similar trend is obtained for other values of $n)$. We observe (see Fig. \ref{fig7}) 
that RES substantially increases with sin$\theta $/$\lambda $: this 
behaviour is partially expected since the high angle intensities are weak, 
but some systematic errors may remain. Luckily the error is smaller for low 
angle intensities: they drive the phasing process to convergence in spite of 
the large residual at high angles. 

\begin{figure}
\includegraphics[width=7cm]{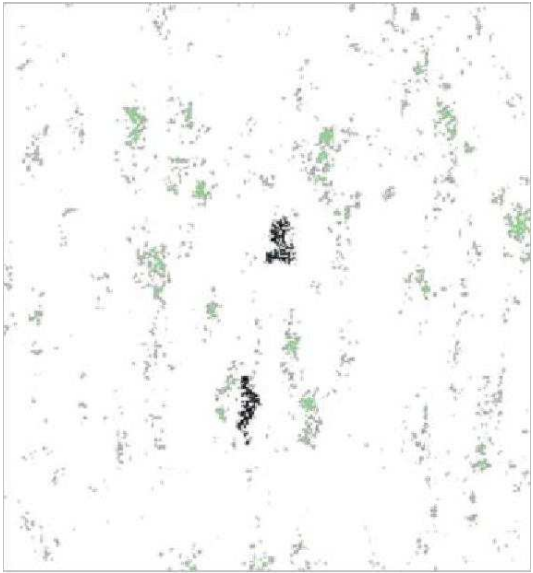}
\caption{Recovered charge density from non-periodic array of 50nm 
diameter gold balls using the HIO algorithm. \label{fig8}}
\end{figure}

In Fig. \ref{fig8} we show the map obtained by application of the ``Shrinkwrap" HIO 
algorithm, in which a support is obtained from the autocorrelation function, 
and then it is improved during HIO iterations. The map was calculated by 
assuming $\vert $F$_{h,k}\vert $ = ($\vert $F$_{h,k}\vert +\vert $ 
F$_{ - h, - k}\vert )$/2 and, for the phases, we assume $\varphi 
_{\mbox{--}h, - k}$= - $\varphi _{h,k}$, ensuring a real object. The map 
correctly reveals the ball positions, but shows a high background level. In 
Table \ref{tab1} we show the values of RES for various $n$, obtained by using the HIO 
algorithm. The RES values are much higher than those obtained by the map 
provided by SIR2002, owing to the fact that the HIO refinement process is 
unable to eliminate the large number of false low intensity peaks. 

\begin{figure}
\includegraphics[width=7cm]{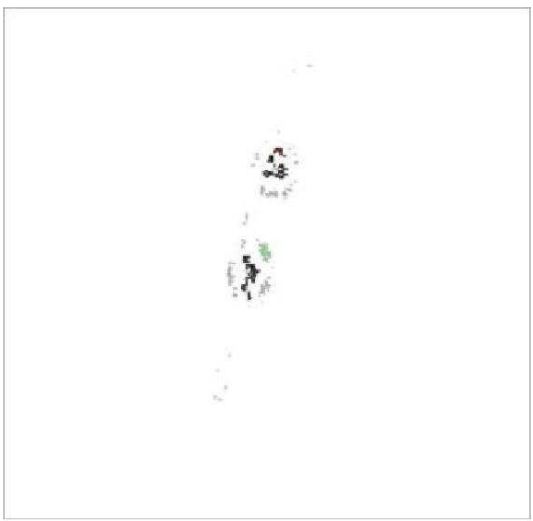}
\caption{Recovered charge density from non-periodic array of 50 nm 
diameter gold balls, using the modified SIR2002 program at the end of STEP 
1. \label{fig9}}
\end{figure}

Finally, it is useful to verify the efficiency of our algorithm when the 
data resolution is low. We therefore stopped the phasing procedure at the 
end of STEP 1, when we had phased NREFTRH = 51193 observed reflections and 
UREFL = 1365 beam-stop reflections. The data resolution was limited to 500 
{\AA}, comparable with the diameter of the balls. SIR2002 algorithm again 
succeeded, and produced the map shown in Fig. \ref{fig9}. We note that: 

\begin{enumerate}
\renewcommand{\theenumi}{\alph{enumi}}
\item
the structural model is essentially correct; 
\item
the RES value is equal to 20, smaller than the value obtained at the end of 
STEP 2.
\end{enumerate}

This last result is of particular interest: it shows that oversampling in 
reciprocal space provides such a high amount of information that the phase 
problem may be solved even in those cases in which the data resolution is 
insufficient for crystal structure solution. 

This conclusion raises a further question: what degree of oversampling is 
optimal for solution of the phase problem? 

In the absence of noise, Shannon's theorem indicates that the Nyquist rate 
allows reconstruction of the diffraction pattern at any angle (between 
samples) if the samples are independent. In the presence of noise, however, 
higher degree of oversampling: a) produces a high correlation between the 
intensities of neighbouring pixels. Thus the total amount of information 
does not increases linearly with the amount of oversampling; b) requires 
small detector pixels, variable working distance, and/or large supports. 
These conditions are not always easily satisfied: the first demands a high 
quality detector, and the second, high spatial coherence in the incident 
beam; c) generates a very large number of reflections (particularly for the 
three-dimensional case), not easily manageable using current computer 
programs. 

\section{Conclusions}

A new phasing algorithm, based on the density modification portion of the 
SIR2002 direct-methods crystallographic program, has been applied to soft 
X-ray scattering from a non-periodic object. The algorithm successfully 
reconstructs unobserved phases from diffuse (continuous) measured scattering 
amplitudes. As a result, images of a cluster of 50 nm diameter gold balls 
have been reconstructed. The results provide an interesting comparison with 
the HIO Fienup-Gerchberg-Saxton algorithm, which operates on closely related 
principles. By contrast, however, no knowledge of the object boundary is 
assumed, and the SIR2002 phasing proceeds as a steadily increasing function 
of spatial frequency.

\appendix
\section

\begin{figure}
\includegraphics[width=7cm]{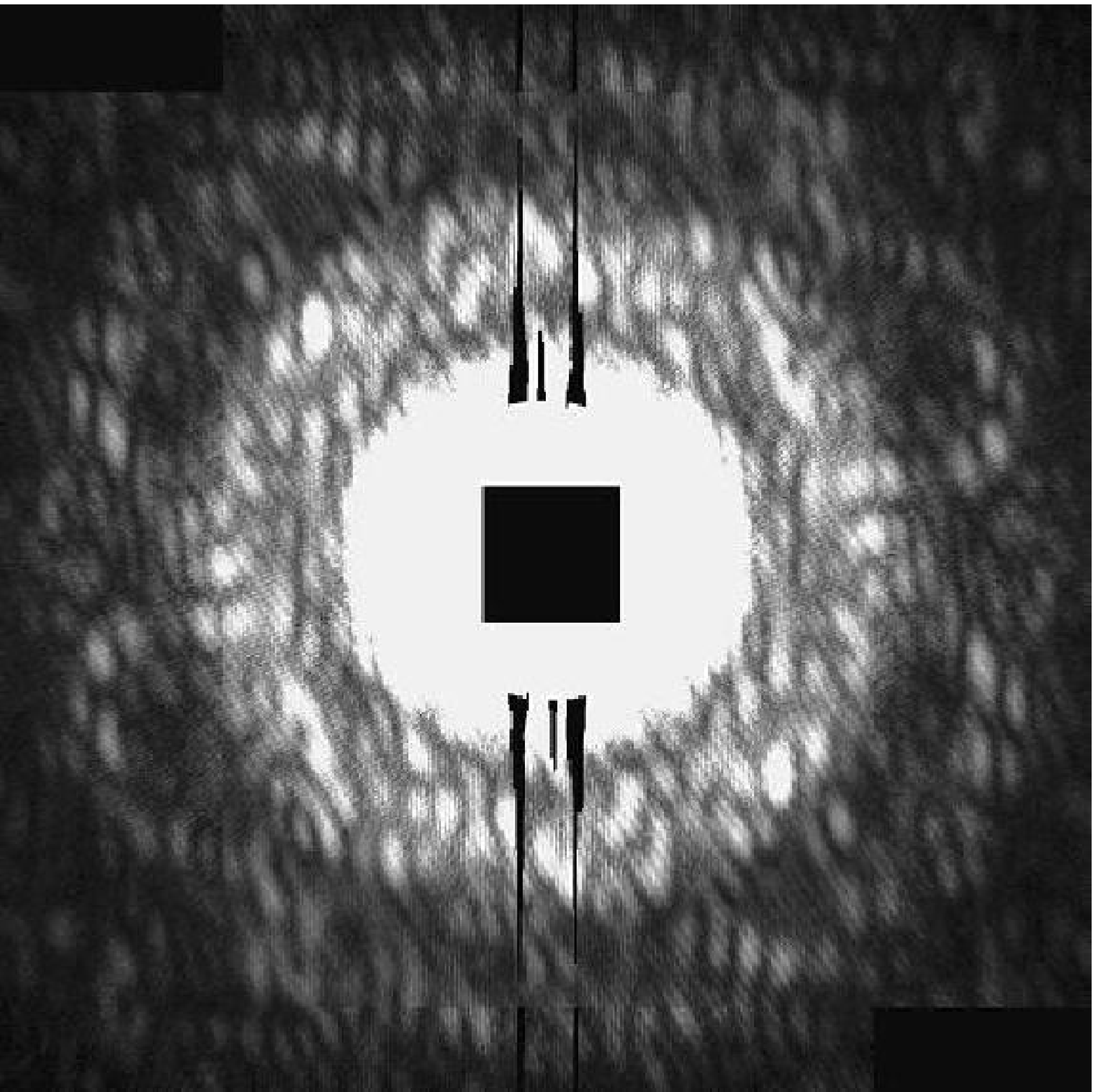}
\vskip -10 pt
\caption{Experimental soft X-ray transmission diffraction pattern 
from the 50nm diameter gold balls shown in Fig.A2. The X-ray wavelength is 
2.5 nm. \label{figa1}}
\end{figure}

\begin{figure}
\includegraphics[width=7cm]{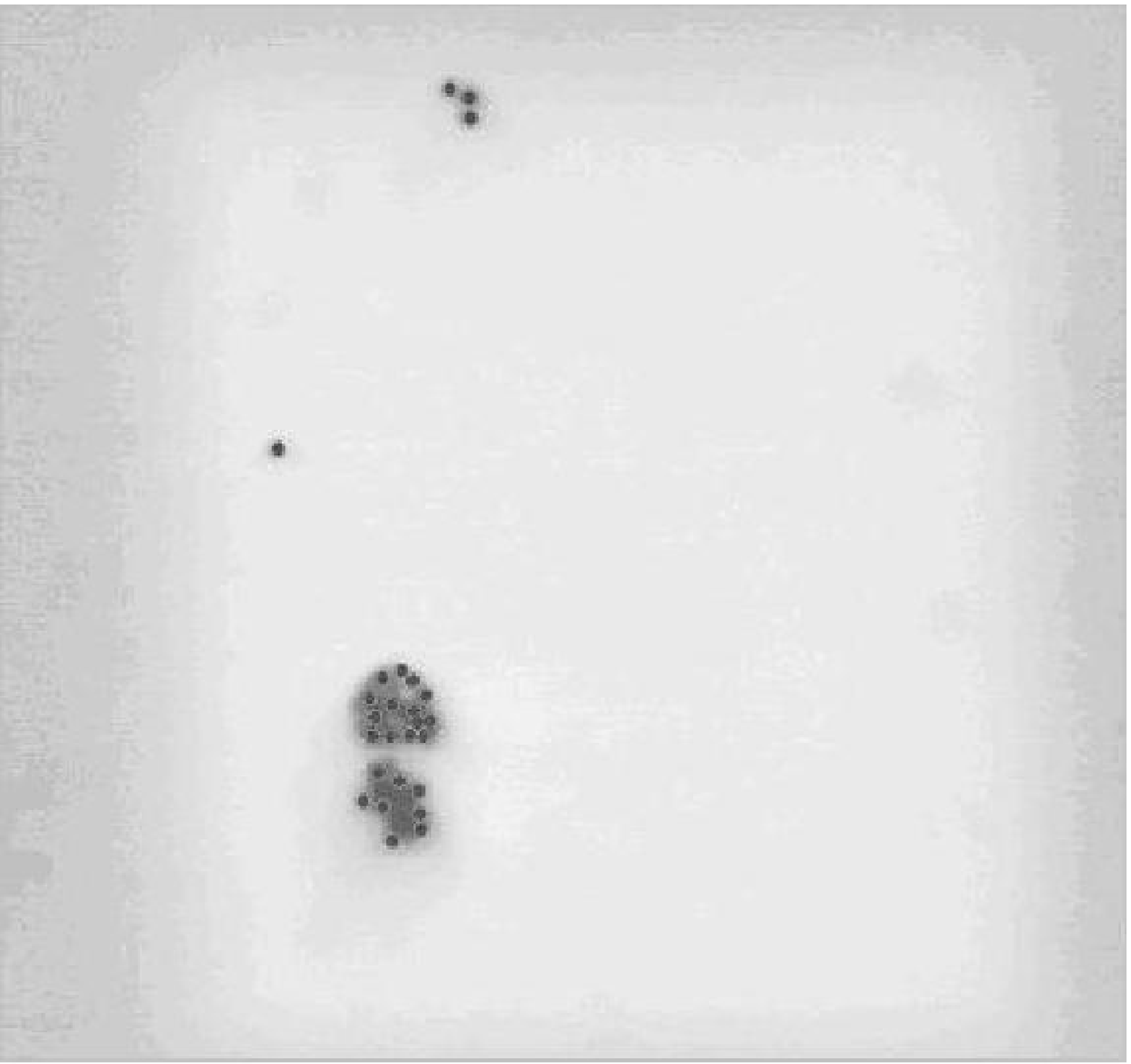}
\vskip -10 pt
\caption{SEM image of a random set of 28 gold balls of 50 nm diameter at 550 eV.\label{figa2}}
\end{figure}

\begin{figure}
\includegraphics[width=7cm]{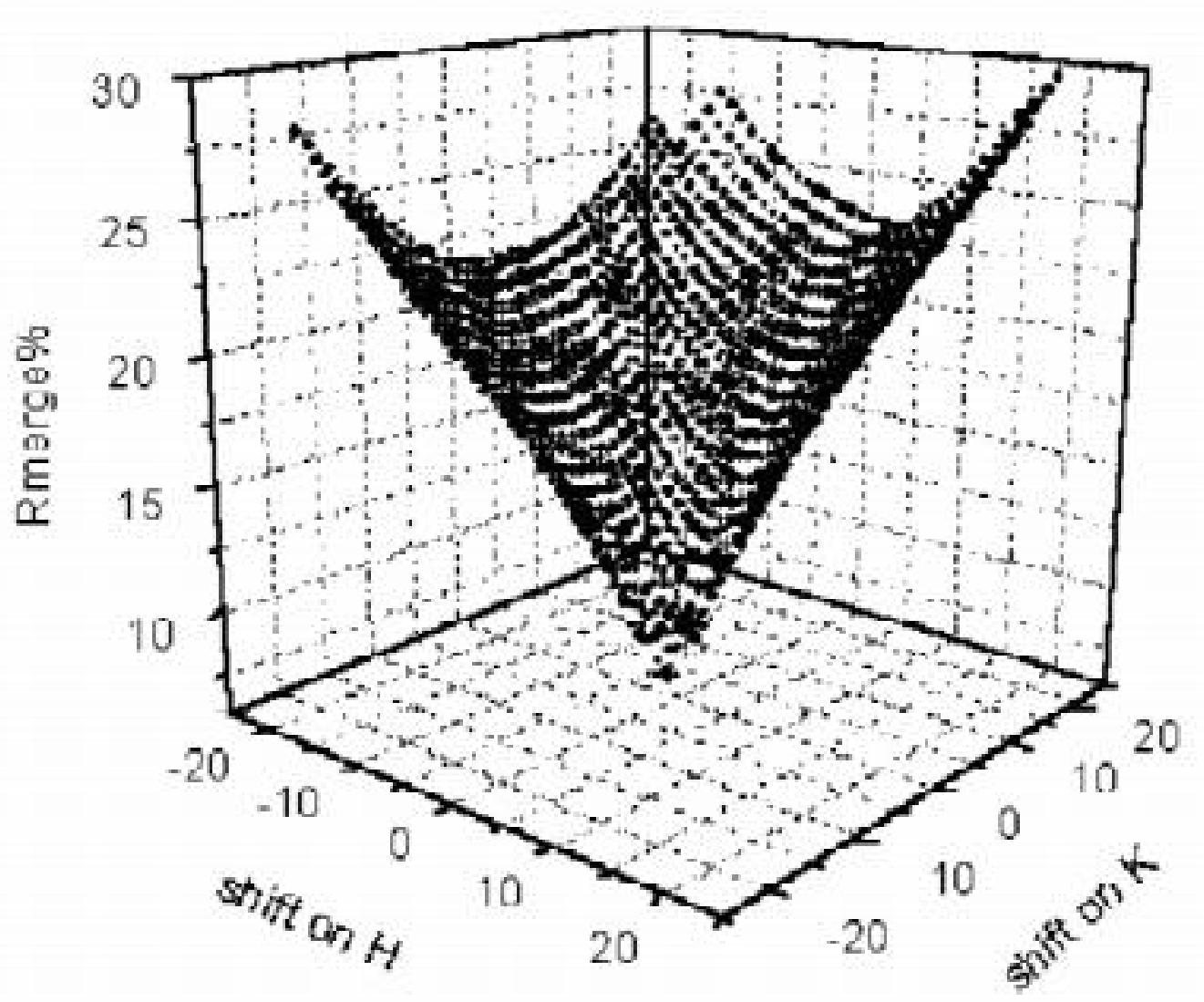}
\vskip -10 pt
\caption{Behaviour of R$_{merge}$ between Friedel opposites as a 
function of the index shift for the gold ball structure shown in Fig. \ref{figa2} \label{figa3}}
\end{figure}

We have calculated R$_{merge}$ as a function of the index shift for the 
pattern shown in Fig \ref{figa1}. This is obtained from a distribution of gold 
balls such as that shown in Fig. \ref{figa2}. The function R$_{merge}$ is shown in 
Fig. \ref{figa3}. The minimum value of the R$_{merge}$ occurs by subtracting 3 to h 
and adding 2 to k. After the origin shift the average discrepancy between 
Friedel amplitudes decreases from 8.99{\%} (obtained for unshifted data) to 
6.33{\%}. The shift for which the minimum is obtained is clearly defined.

\ack{Acknowledgements. }

This work was supported by ARO award DAAD190010500 (JCHS PI) and by the 
National Council of Research (Italy).




\begin{table}
\caption{The RES values for various n  values (see main text ) are shown when the modified SIR2002 and the HIO algorithms are used (column 2 and 3, respectively).}
\begin{tabular}{c c c}
\hline
n& SIR2002	& HIO \\
\hline
1& 23		& 44 \\
2& 23		& 46 \\
3& 23		& 44 \\
4& 23		& 40 \\
5& 24		& 40 \\
6& 25		& 41 \\
7& 27		& 43 \\
\end{tabular}
\label{tab1}
\end{table}


\begin{thebibliography}{}
\bibitem{1}{Bautschke, H., Combettes, P.L. {\&} Luke, D. R. (2002). \emph{J. Opt. Soc. Am}., \textbf{19}, 1344-1345.}
\bibitem{2}{Burla, M.C., Camalli, M., Carrozzini, B., Cascarano, G.L., Giacovazzo, C., 
Polidori, G., Spagna, R. (2003). \emph{J. Appl. Cryst.}, \textbf{36}, 1103.}

\bibitem{3}{De Caro, L., Giacovazzo, C. {\&} Siliqi, D. (2002). \emph{Acta Cryst}., A\textbf{58}, 415 - 423.}
\bibitem{4}{Fienup, J.R. (1982). \emph{Appl. Optics}, \textbf{21}, 2758-2769.}
\bibitem{5}{Fienup, J.R. (1987). \emph{J. Opt. Soc. Am}., \textbf{4}, 118-123.}
\bibitem{6}{He, H., Marchesini, S., Howells, M., Weierstall, U., Hembree, G. {\&} 
Spence, J.C.H. (2003). \emph{Acta Cryst}., A\textbf{59}, 143-152.}

\bibitem{7}{Liao, C., Fiddy, M. {\&} Byrne, C. (1997). \emph{J. Opt. Soc. Am}., \textbf{14}, 3155-3161.}

\bibitem{8}{Marchesini, S., He, H., Chapman, H., Hau-Riege, S., Noy, A., Howells, M., 
Weierstall, U. {\&} Spence, J.C.H. (2003). \emph{Phys Rev}., \textbf{68}, 140101(R) [arXiv:physics.optics/0306174].}
\bibitem{9}{Miao, J., Charalambous, C., Kirz, J. {\&} Sayre, D. (1999). \emph{Nature}, \textbf{400}, 
342-344.}
\bibitem{10}{Miao, J., Ishikawa, T., Johnson, E.H., Lai, B. {\&} Hodgson, K. (2002). 
\emph{Phys. Rev. Lett}., \textbf{89},1-4.}
\bibitem{11}{Paganin, D. {\&} Nugent, K. (1988). \emph{Phys. Rev. Lett}., \textbf{80}, 2586 -- 2589.}
\bibitem{12}{Spence J.C.H., Wu, J.S., Giacovazzo, C., Carrozzini, B., Cascarano, G.L. 
{\&} Padmore, H.A. (2003). \emph{Acta Cryst}., A\textbf{59}, 255-261.}

\bibitem{13}{Stark, H.(1987). \textit{Image recovery: Theory and Applications}. New York: Academic Press.}

\bibitem{14}{Wu, J.S. {\&} Spence, J.C.H. (2003). \emph{Acta Cryst}. A\textbf{59}, 577-583.}

\bibitem{15}{Zuo, J.M., Vartanyants, I.A., Gao, M., Zhang, M. {\&} Nagahara, L.A. (2003). \emph{Science}, \textbf{300}, 1419.}


\end{thebibliography}
\end{document}